\documentclass[10pt]{iopart}
\usepackage{graphicx}
\usepackage{tabls}
\eqnobysec

\newcommand{\beq}{\begin{equation}}
\newcommand{\beqa}{\begin{eqnarray}}
\newcommand{\eeq}{\end{equation}}
\newcommand{\eeqa}{\end{eqnarray}}

\newcommand{\braket}[2]{\langle#1\vert#2\rangle}
\newcommand{\abs}[1]{\vert#1\vert}

\newcommand{\dd}{{\rm d}}
\renewcommand{\det}{\mathop{\rm det}\,}
\newcommand{\etc}{\!\!\!\cdots}
\newcommand{\even}{{({\rm even})}}
\newcommand{\h}{\widehat}
\newcommand{\half}{{\frac{1}{2}}}
\newcommand{\ii}{{\rm i}}
\renewcommand{\o}{\omega}
\newcommand{\odd}{{({\rm odd})}}
\newcommand{\quart}{{\frac{1}{4}}}
\renewcommand{\t}{\widetilde}
\newcommand{\vq}{{\bf q}}
\newcommand{\vx}{{\bf x}}
\newcommand{\vy}{{\bf y}}
\newcommand{\vz}{{\bf z}}
\newcommand{\w}{^{{\rm w}}}
\newcommand{\wo}{\sqrt{\omega}}
\newcommand{\wt}{\widetilde}
\newcommand{\EF}{E_{\rm F}}
\renewcommand{\H}{{\cal H}}

\begin{document}

\title{Return probability of $N$ fermions released from a 1D confining potential}

\author{P L Krapivsky$^{1,2}$, J M Luck$^2$ and K Mallick$^2$}

\address{$^1$ Department of Physics, Boston University, Boston, MA 02215, USA}

\address{$^2$ Institut de Physique Th\'eorique, Universit\'e Paris-Saclay, CEA and CNRS,
91191 Gif-sur-Yvette, France}

\begin{abstract}
We consider $N$ non-interacting fermions prepared in the ground state
of a 1D confining potential and submitted to an instantaneous quench
consisting in releasing the trapping potential.
We show that the quantum return probability
of finding the fermions in their initial state at a later time
falls off as a power law in the long-time regime,
with a universal exponent depending only on $N$
and on whether the free fermions expand over the full line or over a half-line.
In both geometries the amplitudes of this power-law decay
are expressed in terms of finite determinants
of moments of the one-body bound-state wavefunctions in the potential.
These amplitudes are worked out explicitly for the harmonic and square-well potentials.
At large fermion numbers they obey scaling laws
involving the Fermi energy of the initial state.
The use of the Selberg-Mehta integrals stemming from random matrix theory
has been instrumental in the derivation of these results.
\end{abstract}

\eads{\mailto{pkrapivsky@gmail.com},\mailto{jean-marc.luck@ipht.fr},\mailto{kirone.mallick@ipht.fr}}

\maketitle

\section{Introduction}
\label{intro}

Dynamical properties of a quantum system subject to a quench,
i.e., a sudden modification of its Hamiltonian,
have become an active field of research
(see~\cite{CC,caux,fagotti},
and~\cite{Polkovnikov,Eisert,gogolin,dalessio,borgo} for reviews).
This renewed interest has been largely motivated
by experimental progress in cold atom physics,
where the evolution of a condensate
after removing a trapping potential can be scrutinized.
Amongst many possible observables,
the return probability of a quantum system to its initial state
(also referred to as fidelity or quantum Loschmidt echo),
originally proposed by Peres as an indicator of chaos in quantum systems~\cite{Peresbook},
is fully relevant to monitor the dynamics after a quench.
The decay of the quantum return probability with time -- that can intuitively
be related to decoherence -- is found to be of various types:
generically exponential in chaotic systems~\cite{peres},
it can be shown by semi-classical arguments to be slower
in regular systems~\cite{jalabert}.
The situation is more subtle, though,
as various time scales are involved~\cite{manfredi},
and a crossover occurs around the Ehrenfest time (see~\cite{gorin} for a review).
For many-particle systems, decay laws ranging from power-law to super-exponential
have been predicted~\cite{Luna,Campo1,Campo2,Valle,Granot,Stephan2011,Stephan2017}.

In a recent work~\cite{KLM} we studied the quantum return probability
for a system of~$N$ free fermions hopping on a discrete infinite
or semi-infinite lattice, launched from a compact configuration
where the fermions occupy neighboring sites.
We showed that this probability decays algebraically in time,
with an exponent that exhibits an intriguing dependence
on the parity of $N$, due to the combined effects of quantum interferences and discreteness.
We also determined the exact decay amplitudes,
thanks to a mapping to the Selberg-Mehta integrals of random matrix theory~\cite{mehta}.

The aim of the present work is to investigate the return probability of $N$ fermions
after free expansion in a different setting,
namely when the system is prepared in the lowest energy state in a 1D trapping potential.
We first consider the case where the fermions expand over the full line (section~\ref{line}).
At time $t=0$ the confining potential is instantaneously removed.
We show that the return probability falls off as a negative power of time,
with a universal exponent $N^2$.
The associated amplitude is expressed in terms of a finite determinant
of moments of the one-body bound-state wavefunctions in the potential (sections~\ref{gen}
and~\ref{sym}).
It is evaluated exactly in the cases
of harmonic (section~\ref{oh}) and square-well (section~\ref{square}) potentials.
When the fermion number $N$ is large,
the return probability is shown to assume a scaling form
involving the ratio $N/(\EF t)$,
where $\EF$ is the Fermi energy of the initial state (section~\ref{sca}).
All these results are then extended in section~\ref{wline},
with the same setup,
to the situation where the fermions expand only over a semi-infinite line.
The universal decay exponent of the return probability is now $N(2N+1)$.
Section~\ref{disc} contains a brief discussion of our findings.
In~\ref{classical} we give a self-consistent treatment of the problem
of non-colliding classical random walkers,
using an approach that parallels the quantum calculations.
\ref{meh} and~\ref{barn}
are respectively devoted to Mehta integrals and to the Barnes $G$-function.

\section{Non-interacting fermions released on the infinite line}
\label{line}

In this section we consider a system of $N$ non-interacting spinless fermions
on the infinite continuous line.
The system is prepared in the lowest energy state
in an arbitrary confining potential.
At time $t=0$, a quantum quench is performed by
releasing the confining potential.
We are interested in the return probability~$R_N(t)$
that the particles are back to their initial state at a later time $t$,
and especially in the asymptotic decay of this probability in the long-time regime.

\subsection{Generalities}
\label{gen}

The one-body Hamiltonian reads
\beq
\H=\frac{p^2}{2}+V(x),
\eeq
with $p=-\ii\,\dd/\dd x$.
The Planck constant and the fermion mass have been set to unity.
The potential $V(x)$ is confining,
i.e., $V(x)\to+\infty$ as $x\to\pm\infty$.
Let $E_n$ ($n=0,1,\dots$) be the ordered eigenvalues of $\H$
and $\psi_n(x)$ the associated normalized wavefunctions obeying
\beq
-\half\psi_n''(x)+V(x)\psi_n(x)=E_n\psi_n(x).
\eeq

We consider the situation where the system is prepared in its lowest energy state.
The fermions therefore occupy the $N$ lowest bound states ($n=0,\dots,N-1$).
The corresponding many-body wavefunction is the Slater
determinant\footnote{Throughout this work, unless specified explicitly,
determinants are of size $N\times N$,
with indices in the range $1\le m,n\le N$.}
\beq
\braket{\Psi(0)}{\vx}
=\frac{1}{\sqrt{N!}}\det(\psi_{m-1}(x_n)),
\eeq
with $\vx=(x_1,\dots,x_N)$.
At time $t=0$ the confining potential is released,
and so the fermions undergo free expansion over the infinite line.
The return probability reads
\beq
R_N(t)=\abs{A_N(t)}^2,
\label{rdef}
\eeq
with
\beq
A_N(t)=\braket{\Psi(0)}{\Psi(t)}.
\label{adef}
\eeq

The free expansion dynamics is conveniently described in momentum space.
In analogy with the tight-binding case studied in~\cite{KLM}, we have
\beq
A_N(t)=
\int_{-\infty}^\infty\etc\int_{-\infty}^\infty
\prod_{n=1}^N\left(\frac{\dd q_n}{2\pi}\,\e^{-\half\ii tq_n^2}\right)
\abs{\braket{\Psi(0)}{\vq}}^2,
\label{am}
\eeq
\beq
\braket{\Psi(0)}{\vq}
=\frac{1}{\sqrt{N!}}\det(\h\psi_{m-1}(q_n)),
\label{psiq}
\eeq
with $\vq=(q_1,\dots,q_N)$
and
\beq
\h\psi_m(q)=\int_{-\infty}^\infty\psi_m(x)\,\e^{-\ii qx}\,\dd x.
\label{momdef}
\eeq

We are mainly interested in the asymptotic decay of $R_N(t)$.
In the long-time regime, the integral entering~(\ref{am})
is dominated by the region where all the momenta~$q_n$ are small.
In the latter region,
the determinant in~(\ref{psiq}) can be estimated by expanding
each wavefunction in momentum space as
\beq
\h\psi_m(q)=\sum_{k\ge0}M_{m,k}q^k,
\label{psiexp}
\eeq
with
\beq
M_{m,k}=\frac{(-\ii)^k}{k!}\int_{-\infty}^\infty\psi_m(x)\,x^k\,\dd x.
\label{mdef}
\eeq
The leading contribution
is obtained by truncating the expansion~(\ref{psiexp}) at order $k=N-1$.
The array $M$ thus becomes a square matrix of size $N\times N$,
and the sum over its row index can be read as a matrix product.
We thus obtain
\beq
\det(\h\psi_{m-1}(q_n))
\approx\det\left(\sum_{k=1}^NM_{m-1,k-1}q_n^{k-1}\right)
=C_N\Delta_N(\vq),
\label{detexp}
\eeq
where the amplitude
\beq
C_N=\det(M_{m-1,k-1})
\label{cndef}
\eeq
depends on the confining potential,
whereas the universal second factor
\beq
\Delta_N(\vq)=\det(q_n^{m-1})
=\prod_{1\le m<n\le N}(q_n-q_m)
\label{vdm}
\eeq
is the Vandermonde determinant of the momenta.

In the long-time regime,
the expression~(\ref{am}) for the amplitude $A_N(t)$ thus simplifies~to
\beq
A_N(t)\approx\frac{\abs{C_N}^2}{N!}
\int_{-\infty}^\infty\etc\int_{-\infty}^\infty
\prod_{n=1}^N\left(\frac{\dd q_n}{2\pi}\,\e^{-\half\ii tq_n^2}\right)\Delta_N^2(\vq).
\eeq
The key observation is that the above integral is the analytical continuation to $a=\half\ii t$
of the Mehta integral~(\ref{meh1q}).
We thus obtain the following prediction for the asymptotic decay
of the quantum return probability in the long-time regime:
\beq
R_N(t)\approx\frac{\abs{C_N}^4G(N+1)^2}{(2\pi)^Nt^{N^2}},
\label{ras}
\eeq
where $G$ denotes the Barnes $G$-function (see~\ref{barn}).

The universal decay exponent~$N^2$
can be recovered by the following heuristic argument~\cite{KLM}.
The wavefunction of $N$ non-interacting fermions can be estimated by expressing that
the particles spread on a ballistic scale $L(t)\sim t$ and that the wavefunction
vanishes when two particles occupy the same position.
Thus one can write
\beq
\abs{\Psi(\vx,t)}\sim K_N(t)\prod_{1\le m < n\le N}\abs{x_m - x_n}
\eeq
if $\abs{x_i}\le L(t)$, whereas $\Psi$ essentially vanishes otherwise.
Dimensional analysis implies that the normalization scales as
$K_N(t)\sim t^{-N^2/2}$,
whence the exponent $N^2$ derived above by more quantitative means
for the return probability.

The expression~(\ref{ras})
only depends on the confining potential $V(x)$
through the prefactor~$C_N$, given by~(\ref{cndef}).
Hereafter we shall calculate $C_N$ exactly for harmonic and square-well potentials
in sections~\ref{oh} and~\ref{square}.
We shall also estimate its large-$N$ asymptotics
for a general confining potential $V(x)$ in section~\ref{sca},
using a heuristic semi-classical analysis.

For a single particle,~(\ref{ras}) reads
\beq
R_1(t)\approx\frac{\abs{M_{00}}^4}{2\pi t},
\eeq
where
\beq
M_{00}=\int_{-\infty}^\infty\psi_0(x)\,\dd x
\eeq
is the integral of the ground-state wavefunction.
This quantity can be used to define the spatial extent $\ell$
of the ground state by setting $\abs{M_{00}}=\sqrt{\ell}$.
The resulting expression,
\beq
R_1(t)\approx\frac{\ell^2}{2\pi t},
\label{r1}
\eeq
can be read as $R_1(t)\sim\ell/L(t)$,
where the dynamical length $L(t)\sim t/\ell$
represents the ballistic spreading of the quantum particle,
whose momentum scale $p\sim1/\ell$ is dictated by the uncertainty principle.
An alternative interpretation of~(\ref{r1})
is that it exhibits the formal diffusive scaling of the Schr\"odinger equation.

\subsection{Considerations about symmetry}
\label{sym}

In the situation where the confining potential is symmetric, i.e., $V(-x)=V(x)$,
the eigenstates have a definite parity, i.e.,
\beq
\psi_n(-x)=(-1)^n\psi_n(x),
\eeq
and so the matrix element $M_{m,k}$ vanishes if $m+k$ is odd.
The expression~(\ref{cndef}) of~$C_N$ therefore splits into the product of two determinants
corresponding to each parity sector, namely
\beqa
C_{2p}&=&c_p^\even\,c_p^\odd,
\label{cce}
\\
C_{2p+1}&=&c_{p+1}^\even\,c_p^\odd,
\label{cco}
\eeqa
with
\beqa
c_p^\even&=&\det(M_{2k,2l})_{0\le k,l\le p-1},
\label{cedef}
\\
c_p^\odd&=&\det(M_{2k-1,2l-1})_{1\le k,l\le p}.
\label{codef}
\eeqa

Let us illustrate this in the case of two fermions.
For an arbitrary potential,
we have $C_2=M_{00}M_{11}-M_{10}M_{01}$,
and so
\beq
R_2(t)\approx\frac{\abs{M_{00}M_{11}-M_{10}M_{01}}^4}{4\pi^2t^4}.
\eeq
For a symmetric potential,
we have $c_1^\even=M_{00}$, $c_1^\odd=M_{11}$, hence $C_2=M_{00}M_{11}$, and so
\beq
R_2(t)\approx\frac{\abs{M_{00}M_{11}}^4}{4\pi^2t^4}.
\eeq

\subsection{Harmonic potential}
\label{oh}

In this section we consider the case of a harmonic potential well:
\beq
V(x)=\frac{\o^2x^2}{2}.
\eeq
The quantum harmonic oscillator is a textbook example of an exactly solvable system.
This is seemingly the only example
where the full time dependence of the return probability can be worked out explicitly.

The energy levels read
\beq
E_n=\left(n+\half\right)\o\qquad(n=0,1,\dots).
\eeq
The corresponding wavefunctions have essentially the same form
in position space and in momentum space, namely
\beqa
\psi_n(x)&=&\frac{(\o/\pi)^{1/4}}{\sqrt{2^nn!}}\,H_n(x\wo)\,\e^{-\o x^2/2},
\\
\h\psi_n(q)&=&(-\ii)^n\frac{(4\pi/\o)^{1/4}}{\sqrt{2^nn!}}\,H_n(q/\wo)\,\e^{-q^2/(2\o)},
\label{hpsi}
\eeqa
where
\beq
H_n(z)=\sum_{k=0}^{\lfloor{n/2}\rfloor}(-1)^k\frac{n!}{k!(n-2k)!}(2z)^{n-2k}
\eeq
are the Hermite polynomials~\cite{Abramowitz}.
Using~(\ref{hpsi}), and introducing the notation $z_n=q_n/\sqrt{\o}$,
the expression~(\ref{psiq}) can be recast as
\beqa
\braket{\Psi(0)}{\vq}&=&
\left(-\frac{\ii}{\sqrt{2}}\right)^{N(N-1)/2}
\frac{(4\pi/\o)^{N/4}}{\sqrt{G(N+2)}}
\nonumber\\
&\times&\exp\left(-\half\sum_{n=1}^Nz_n^2\right)
\,\det(H_{m-1}(z_n)).
\eeqa
The latter determinant can be evaluated as follows.
By subtracting from the $m$th line a suitably chosen linear combination of the previous ones,
the Hermite polynomial $H_{m-1}(z_n)$ can be replaced by its leading term $(2z_n)^{m-1}$.
The determinant is therefore proportional to a Vandermonde determinant:
\beq
\det(H_{m-1}(z_n))=2^{N(N-1)/2}\Delta_N(\vz),
\label{hvdm}
\eeq
with $\vz=(z_1,\dots,z_N)$.

The expression~(\ref{am}) of the amplitude $A_N(t)$ therefore reads
\beqa
A_N(t)&=&\frac{2^{N(N-1)/2}}{{\pi^{N/2}}G(N+2)}
\nonumber\\
&\times&\int_{-\infty}^\infty\etc\int_{-\infty}^\infty
\prod_{n=1}^N\left(\dd z_n\,\e^{-(1+\half\ii\o t)z_n^2}\right)\Delta_N^2(\vz).
\eeqa
The above integral is a Mehta integral of the form~(\ref{meh1q}) at any finite time $t$.
This is a very peculiar feature of the harmonic oscillator.
We thus obtain the remarkably simple exact expressions\footnote{The
normalization $A_N(0)=1$ for all $N$ provides a useful check of the formalism.}
\beqa
A_N(t)&=&\left(1+\frac{\ii\o t}{2}\right)^{-N^2/2},
\\
R_N(t)&=&\left(1+\frac{\o^2 t^2}{4}\right)^{-N^2/2},
\eeqa
which hold for all fermion numbers $N$ and all times $t$.
The return probability therefore exhibits the power-law decay
\beq
R_N(t)\approx\left(\frac{2}{\o t}\right)^{N^2},
\label{roh}
\eeq
with exponent $N^2$, in agreement with~(\ref{ras}), and a simple prefactor.

Anticipating the analysis of the large-$N$ regime presented in section~\ref{sca},
we recast the above formula in terms of the Fermi energy $\EF$,
defined as the energy of the last occupied one-particle state.
In the case of the harmonic oscillator, we have $\EF\approx N\o$, hence
\beq
R_N(t)\sim\left(\frac{2N}{\EF t}\right)^{N^2}.
\label{rasoh}
\eeq

\subsection{Square-well potential}
\label{square}

In this section we consider the case of a square-well potential:
\beq
V(x)=\left\{\matrix{
0\hfill&(\abs{x}<L/2),\cr
+\infty\quad&(\abs{x}>L/2).
}\right.
\eeq
The particles are thus confined between two impenetrable walls at $x=\pm L/2$.
This is another textbook example of an exactly solvable system.
The energy levels read
\beq
E_n=(n+1)^2\frac{\pi^2}{2L^2}\qquad(n=0,1,\dots).
\eeq
The corresponding wavefunctions are as follows.

\noindent $\bullet$
Even sector ($n=2p$, $p=0,1,\dots$):
\beqa
\psi_{2p}(x)&=&\sqrt{\frac{2}{L}}\,\cos\frac{(2p+1)\pi x}{L},
\\
\h\psi_{2p}(q)&=&(-1)^p\,\pi\sqrt{2L}\,\frac{2(2p+1)}{(2p+1)^2\pi^2-q^2L^2}\,\cos\frac{qL}{2}.
\label{psieven}
\eeqa

\noindent $\bullet$
Odd sector ($n=2p-1$, $p=1,2,\dots$):
\beqa
\psi_{2p-1}(x)&=&\sqrt{\frac{2}{L}}\,\sin\frac{2p\pi x}{L},
\\
\h\psi_{2p-1}(q)&=&(-1)^p\,\ii\pi\sqrt{2L}\,\frac{4p}{4p^2\pi^2-q^2L^2}\,\sin\frac{qL}{2}.
\label{psiodd}
\eeqa

At variance with the case of the harmonic oscillator,
the return probability cannot be evaluated exactly at finite time.

The prefactor $C_N$ entering the asymptotic decay law~(\ref{ras}) of $R_N(t)$
can however be evaluated exactly, for all values of the fermion number $N$.
The key point of the derivation resides in the following observation.
The determinants $c_p^\even$ and $c_p^\odd$, introduced in~(\ref{cedef}),~(\ref{codef}),
can be simplified along the lines of the derivation of~(\ref{hvdm}).
In the even sector,
it is legitimate to replace $\cos(qL/2)$ by unity in the expression~(\ref{psieven})
of $\h\psi_{2p}(q)$,
as this amounts to subtracting from the rows of the matrix $M_{2k,2l}$
a suitably chosen linear combination of the previous ones.
We denote by $\t M$ the matrix simplified in this way.
Similarly, for the odd sector,
$\sin(qL/2)$ can be linearized to $qL/2$
in the expression~(\ref{psiodd}) of $\h\psi_{2p-1}(q)$.

In the even sector, we thus obtain
\beqa
&&c_p^\even=\det(\t M_{2k,2l})_{0\le k,l\le p-1},
\nonumber\\
&&\t M_{2k,2l}=\frac{(-1)^k\,2\sqrt{2}}{((2k+1)\pi)^{2l+1}}\,L^{2l+1/2},
\eeqa
and so
\beq
c_p^\even=\frac{2^{3p/2}}{\pi^{p^2}(2p-1)!!}\,d_p^\even\,L^{p(p-1/2)},
\eeq
with
\beq
d_p^\even=(-1)^{p(p-1)/2}\det((2k+1)^{-2l})_{0\le k,l\le p-1}.
\label{dpedef}
\eeq
In the odd sector, we have
\beqa
&&c_p^\odd=\det(\t M_{2k-1,2l-1})_{1\le k,l\le p},
\nonumber\\
&&\t M_{2k-1,2l-1}=\frac{(-1)^k\,\ii\sqrt{2}}{(2k\pi)^{2l-1}}\,L^{2l-1/2},
\eeqa
and so
\beq
c_p^\odd=\frac{(-\ii)^p\,2^{p/2}}{(2\pi)^{p^2}p!}\,d_p^\odd\,L^{p(p+1/2)},
\eeq
with
\beq
d_p^\odd=(-1)^{p(p-1)/2}\det(k^{-2(l-1)})_{1\le k,l\le p}.
\label{dpodef}
\eeq
The determinants $d_p^\even$ and $d_p^\odd$ are evaluated in~\ref{barn}.
Their explicit expressions~(\ref{dpe}) and~(\ref{dpo}) yield
\beqa
c_p^\even&=&\frac{2^{3p^2}}{\pi^{p^2}}\left(\frac{p!}{(2p)!}\right)^{2p}\sqrt{\frac{G(2p+2)}{p!}}\,L^{p(p-1/2)},
\\
c_p^\odd&=&\frac{(-\ii)^p}{(2\pi)^{p^2}\,p!^{2p}}\sqrt{\frac{G(2p+2)}{p!}}\,L^{p(p+1/2)}.
\label{cores}
\eeqa
Inserting the above results into~(\ref{cce}),~(\ref{cco}), we get
\beq
\abs{C_N}
=\frac{G(N+2)}{N!^N\Gamma(\frac{N}{2}+1)}\left(\frac{2L}{\pi}\right)^{N^2/2},
\eeq
irrespective of the parity of $N$.

We thus obtain the following asymptotic decay law for the return probability
\beq
R_N(t)\approx K_N\left(\frac{L^2}{\pi^2t}\right)^{N^2},
\label{rsquare}
\eeq
where the exponent $N^2$ is in agreement with~(\ref{ras}),
and the prefactor reads
\beq
K_N=\frac{2^{2N^2}G(N+1)^6}{(2\pi)^NN!^{4(N-1)}\Gamma(\frac{N}{2}+1)^4}.
\eeq
In particular
\beq
K_1=\frac{32}{\pi^3},\quad
K_2=\frac{4}{\pi^2},\quad
K_3=\frac{2^{21}}{3^{12}\pi^5}.
\eeq

When the fermion number $N$ becomes large,
the leading behavior of $K_N$
can be derived by using the asymptotic expansions~(\ref{lnf}) and~(\ref{lnsf}).
Keeping only terms in~$N^2$, we obtain
\beq
\ln K_N\approx -N^2\left(\ln N-2\ln 2+\half\right).
\eeq
The expression~(\ref{rsquare}) therefore simplifies to
\beq
R_N(t)\sim\left(\frac{4L^2}{\e^{1/2}\pi^2 Nt}\right)^{N^2}.
\label{rnsquare}
\eeq

Anticipating again the analysis of section~\ref{sca},
we recast the above formula in terms of the Fermi energy $\EF$.
For a square-well potential, the Fermi energy grows as
$\EF\approx N^2\pi^2/(2L^2)$ and the above formula can be rewritten as
\beq
R_N(t)\sim\left(\frac{2\e^{-1/2}\,N}{\EF t}\right)^{N^2}.
\label{rassquare}
\eeq

\subsection{Scaling at large $N$}
\label{sca}

We now focus our attention onto the regime where the fermion number $N$ is large.
By observing the formulas~(\ref{rasoh}) and~(\ref{rassquare})
found for the harmonic and the square-well potential,
it is tempting to propose
the following scaling Ansatz for the return probability
for an arbitrary confining potential $V(x)$:
\beq
R_N(t)\sim\left(\frac{BN}{\EF t}\right)^{N^2}.
\label{rnas}
\eeq
This scaling law is meant to hold in the regime
where the long-time limit is taken before the limit of a large fermion number.
In the denominator, $\EF$ is the Fermi energy,
i.e., the energy of the last occupied one-particle state.
The occurrence of the dimensionless combination $\EF t$ is quite natural.
In the numerator, $B$ appears as a numerical constant of order unity,
which depends on the confining potential $V(x)$.
The above results give $B=2$ for the harmonic oscillator
and $B=2\e^{-1/2}=1.213061$ for the square-well potential.

The Ansatz~(\ref{rnas}) is corroborated by the following heuristic semi-classical analysis
for a general symmetric power-law confining potential,
\beq
V(x)=g\abs{x}^a,
\eeq
with arbitrary growth exponent $a>0$.
Consider a highly excited bound state in this potential, with energy $E_n$ ($n\gg1$).
The wavefunction $\psi_n(x)$ exhibits two turning points at $x=\pm x_n$,
such that $E_n=gx_n^a$.
The energy $E_n$ is given by the semi-classical quantization formula~\cite{Bender}
\beq
2\int_0^{x_n}\sqrt{2(E_n-gx^a)}\,\dd x\approx\left(n+\half\right)\pi.
\eeq
Some algebra leads to
\beq
x_n\approx\left(\frac{\lambda n}{\sqrt{g}}\right)^\beta,
\qquad
E_n\approx g^\beta(\lambda n)^{2(1-\beta)},
\label{xeturn}
\eeq
with
\beq
\beta=\frac{2}{a+2},\qquad
\lambda=\sqrt{\frac{\pi}{2}}\,
\frac{\Gamma\!\left(\frac{3}{2}+\frac{1}{a}\right)}{\Gamma\!\left(1+\frac{1}{a}\right)}.
\eeq
We have therefore $\EF\approx g^\beta(\lambda N)^{2(1-\beta)}$.
The wavefunction $\psi_n(x)$ is rapidly oscillating in the allowed region ($\abs{x}<x_n$),
and exponentially decaying in the forbidden regions ($\abs{x}>x_n$).
Its amplitude is maximal in the transition regions near the turning points.
As a consequence, for large values of the integers $m$ and $n$,
the integral entering the expression~(\ref{mdef}) of $M_{m,k}$
can be expected to be dominated by the vicinity of the turning points.
This heuristic argument yields the rough estimate
\beq
M_{m,k}\sim\frac{x_m^k}{k!};
\label{mturn}
\eeq
and hence
\beq
\ln\abs{M_{m,k}}\approx k\left(\beta\ln\frac{\lambda m}{\sqrt{g}}-\ln k+1\right).
\eeq
The leading scaling behavior of $C_N$ can be read off from the above estimate
by replacing $m$ by $N$ in the argument of the first logarithm
and summing the resulting expression over $k$.
This yields
\beq
\ln\abs{C_N}\approx\frac{N^2}{2}\left((\beta-1)\ln N+\beta\ln\frac{\lambda}{\sqrt{g}}+\cdots\right).
\eeq
Inserting this estimate into~(\ref{ras}),
we find that the return probability indeed obeys the scaling law~(\ref{rnas}).
The case of a square-well potential is recovered in the $\beta\to0$ limit.
The above line of reasoning is however too crude to predict the numerical constant $B$.

\section{Non-interacting fermions released on the semi-infinite line}
\label{wline}

We now consider the same problem on the semi-infinite line ($x>0$).
We assume that there is an impenetrable wall at the origin.
The system is prepared in the lowest energy state
in the presence of an arbitrary confining potential $V\w(x)$ acting for $x>0$.\footnote{The
superscript `w' reminds of the permanent presence of an impenetrable wall at the origin.}
At time $t=0$ the confining potential is released but the wall at the origin is kept,
so that the particles expand over the semi-infinite line.
We are again interested in the return probability $R\w_N(t)$,
and especially in its asymptotic decay.

\subsection{Generalities}
\label{wgen}

The one-body Hamiltonian reads
\beq
\H\w=\frac{p^2}{2}+V\w(x)\qquad(x>0),
\eeq
with Dirichlet boundary condition at the origin.
Let $E\w_n$ ($n=0,1,\dots$) be the ordered eigenvalues of $\H\w$
and $\psi\w_n(x)$ the associated normalized wavefunctions.

At time $t=0$ the confining potential is released,
and so the fermions undergo free expansion over the semi-infinite line.
The return probability reads
\beq
R\w_N(t)=\abs{A\w_N(t)}^2,
\label{wrdef}
\eeq
with
\beq
A\w_N(t)=\braket{\Psi\w(0)}{\Psi\w(t)}.
\label{wadef}
\eeq

The free expansion dynamics is again best described in momentum space.
We~have
\beq
A\w_N(t)=
\int_0^\infty\etc\int_0^\infty
\prod_{n=1}^N\left(\frac{\dd q_n}{\pi}\,\e^{-\half\ii tq_n^2}\right)
\abs{\braket{\Psi\w(0)}{\vq}}^2,
\label{wam}
\eeq
with
\beq
\braket{\Psi\w(0)}{\vq}
=\frac{1}{\sqrt{N!}}\det(\h\psi\w_{m-1}(q_n))
\label{wpsiq}
\eeq
and
\beq
\h\psi\w_m(q)=\sqrt{2}\int_0^\infty\psi\w_m(x)\,\sin qx\,\dd x.
\label{wmomdef}
\eeq

In the long-time regime, the integral entering~(\ref{wam})
is again dominated by the region where all the momenta $q_n$ are small.
In the latter region,
the determinant in~(\ref{wpsiq}) can be estimated by expanding
each wavefunction in momentum space~as
\beq
\h\psi\w_m(q)=\sqrt{2}\sum_{k\ge0}M\w_{m,k}q^{2k+1},
\label{wpsiexp}
\eeq
with
\beq
M\w_{m,k}=\frac{(-1)^k}{(2k+1)!}\int_0^\infty\psi\w_m(x)\,x^{2k+1}\,\dd x.
\label{wmdef}
\eeq
The leading contribution
is again obtained by truncating the expansion~(\ref{wpsiexp}) at order $k=N-1$.
We thus obtain
\beq
\det(\h\psi\w_{m-1}(q_n))
\approx 2^{N/2}\,C\w_N\,\prod_{n=1}^N q_n\,\Delta_N(\vq^2),
\label{wdetexp}
\eeq
where
\beq
C\w_N=\det(M\w_{m-1,k-1}),
\label{wcndef}
\eeq
and $\vq^2=(q_1^2,\dots,q_N^2)$.
In the long-time regime,
the expression~(\ref{wadef}) for the amplitude $A\w_N(t)$ thus simplifies to
\beq
A\w_N(t)\approx\frac{2^N\abs{C\w_N}^2}{N!}
\int_0^\infty\etc\int_0^\infty
\prod_{n=1}^N\left(\frac{\dd q_n}{\pi}\,q_n^2\,\e^{-\half\ii tq_n^2}\right)\Delta_N^2(\vq^2).
\eeq
The above integral is proportional to the analytical continuation to $a=\half\ii t$
of the Mehta integral~(\ref{meh2q}).
We thus obtain the following asymptotic prediction for the return probability
in the long-time regime:
\beq
R\w_N(t)\approx\frac{\abs{C\w_N}^4G(2N+2)}{\pi^N N!\,t^{N(2N+1)}}.
\label{wras}
\eeq
The return probability again exhibits a universal power-law decay, with exponent $N(2N+1)$.
This exponent can also be predicted by means of a heuristic argument (see~\cite{KLM}).
The dependence of the above prediction on the confining potential $V\w(x)$
is entirely contained in the prefactor $C\w_N$, given by~(\ref{wcndef}).

For one single particle,~(\ref{wras}) reads
\beq
R\w_1(t)\approx\frac{2\abs{M\w_{00}}^4}{\pi t^3},
\eeq
with
\beq
M\w_{00}=\int_0^\infty x\,\psi\w_0(x)\,\dd x.
\eeq
This integral can be used to define the spatial extent $\ell\w$
of the ground state by setting $\abs{M\w_{00}}=\left(\ell\w\right)^{3/2}$.
The resulting expression,
\beq
R_1(t)\approx\frac{2\left(\ell\w\right)^6}{\pi t^3},
\eeq
again reflects the formal diffusive scaling of the Schr\"odinger equation.

\subsection{Considerations about symmetry}
\label{wsym}

It is worth comparing the one-sided situation,
i.e., free expansion on the semi-infinite line after preparation in the confining
potential $V\w(x)$ for $x>0$,
with the two-sided situation (section~\ref{line}) with potential
\beq
V(x)=V\w(\abs{x}),
\label{vsym}
\eeq
obtained by symmetrizing $V\w(x)$.

The correspondence between both situations goes as follows.
The Dirichlet boundary condition at the origin
implies that the wavefunctions $\psi\w_n(x)$ essentially coincide with
the odd wavefunctions of the two-sided problem, namely
\beq
\psi\w_n(x)=\sqrt{2}\,\psi_{2n+1}(x)
\label{corr}
\eeq
for $x>0$ and $n=0,1,\dots$,
where the factor $\sqrt{2}$ ensures the correct normalizations.
In momentum space,
with the definitions~(\ref{momdef}) and~(\ref{wmomdef}), this reads
\beq
\h\psi\w_n(q)=\ii\,\h\psi_{2n+1}(q).
\label{psicor}
\eeq
We have therefore
\beq
M\w_{m,k}=\frac{\ii}{\sqrt{2}}\,M_{2m+1,2k+1}.
\eeq
In particular
\beq
C\w_N=\left(\frac{\ii}{\sqrt{2}}\right)^Nc_N^\odd.
\label{cforco}
\eeq

\subsection{Harmonic potential}
\label{woh}

In this section we consider the case of a harmonic half-well:
\beq
V\w(x)=\frac{\o^2x^2}{2}\qquad(x>0).
\eeq
This will again be the only example
where the full time dependence of the return probability $R\w_N(t)$
can be worked out explicitly.

We shall exploit the correspondence underlined in section~\ref{wsym}.
The two-sided situation is that of a harmonic well, studied in section~\ref{oh}.
Using~(\ref{psicor}), as well as~(\ref{hpsi}) and~(\ref{pid}),
and introducing the notation $z_n=q_n/\sqrt{\o}$,
the expression~(\ref{wpsiq}) can be recast as
\beqa
\braket{\Psi\w(0)}{\vq}&=&
\left(-\half\right)^{N(N-1)/2}
\left(\frac{(2\pi/\o)^N}{N!G(2N+2)}\right)^{1/4}
\nonumber\\
&\times&\exp\left(-\half\sum_{n=1}^Nz_n^2\right)
\,\det(H_{2m-1}(z_n)).
\eeqa
The latter determinant can be simplified along the lines of the derivation of~(\ref{hvdm}).
We thus obtain
\beq
\det(H_{2m-1}(z_n))=2^{N^2}\prod_{i=1}^Nz_n\,\Delta_N(\vz^2),
\label{whvdm}
\eeq
with $\vz^2=(z_1^2,\dots,z_N^2)$.

The expression~(\ref{wam}) of the amplitude $A\w_N(t)$ therefore reads
\beqa
A\w_N(t)&=&\frac{2^{N(2N+1)/2}}{\sqrt{\pi^NN!\,G(2N+2)}}
\nonumber\\
&\times&\int_{-\infty}^\infty\etc\int_{-\infty}^\infty
\prod_{n=1}^N\left(\dd z_n\,z_n^2\,\e^{-(1+\half\ii\o t)z_n^2}\right)\Delta_N^2(\vz^2).
\eeqa
The above integral is a Mehta integral of the form~(\ref{meh2q}) at any finite time $t$.
We again obtain the remarkably simple exact expressions\footnote{The
normalization $A\w_N(0)=1$ again provides a useful check of the formalism.}
\beqa
A\w_N(t)&=&\left(1+\frac{\ii\o t}{2}\right)^{-N(2N+1)/2},
\\
R\w_N(t)&=&\left(1+\frac{\o^2 t^2}{4}\right)^{-N(2N+1)/2},
\eeqa
for all fermion numbers $N$ and all times $t$.
The return probability therefore exhibits the power-law decay
\beq
R\w_N(t)\approx\left(\frac{2}{\o t}\right)^{N(2N+1)},
\label{wroh}
\eeq
with exponent $N(2N+1)$, in agreement with~(\ref{wras}),
and a simple prefactor.

\subsection{Square-well potential}
\label{wsquare}

In this section we consider the case of the square-well potential:
\beq
V(x)=\left\{\matrix{
0\hfill&(0<x<L/2),\cr
+\infty\quad&\mbox{else}.\hfill
}\right.
\eeq
The particles are thus confined between two impenetrable walls at $x=0$ and $x=L/2$.
The right wall is removed instantaneously at time $t=0$,
whereas the left one is maintained permanently.
This setting can be viewed as a special case of the problem studied
long ago by Doescher and Rice~\cite{DR},
namely a square well with a wall moving at constant velocity.
The present problem of an instantaneous release corresponds to the infinite velocity limit.

We shall again make use of the correspondence underlined in section~\ref{wsym}.
The two-sided situation is that considered in section~\ref{square}.
Inserting the expression~(\ref{cores}) of $c_N^\odd$ into~(\ref{cforco}),
we readily obtain
\beq
C\w_N=\frac{1}{(2\pi)^{N^2}N!^{2N}}\sqrt{\frac{G(2N+2)}{2^N N!}}\,L^{N(N+1/2)},
\eeq
and finally
\beq
R\w_N(t)\approx K\w_N\left(\frac{L^2}{4\pi^2t}\right)^{N(2N+1)},
\label{wrsquare}
\eeq
where the exponent $N(2N+1)$ is again in agreement with~(\ref{wras}),
and the prefactor reads
\beq
K\w_N=\frac{\pi^NG(2N+2)^3}{N!^{8N+3}}.
\eeq
In particular
\beq
K\w_1=8\pi,\quad
K\w_2=\frac{3^6\pi^2}{2^4},\quad
K\w_3=\frac{2^95^6\pi^3}{3^{12}}.
\eeq

When the fermion number $N$ becomes large,
the leading decay law of $K\w_N$
can again be derived by using the asymptotic expansions~(\ref{lnf}) and~(\ref{lnsf}).
Keeping only terms in $N^2$, we obtain
\beq
\ln K\w_N\approx-2N^2\left(\ln N-3\ln 2+\half\right).
\eeq
The expression~(\ref{wrsquare}) therefore simplifies to
\beq
R\w_N(t)\sim\left(\frac{2L^2}{\e^{1/2}\pi^2 Nt}\right)^{N(2N+1)}.
\label{wrnsquare}
\eeq

\subsection{Scaling at large $N$}
\label{wsca}

In the regime where the fermion number $N$ is large,
along the lines of section~\ref{sca},
we propose the following scaling Ansatz for the return probability:
\beq
R\w_N(t)\sim\left(\frac{B\w N}{\EF\w t}\right)^{N(2N+1)}.
\label{wrnas}
\eeq

Let us again begin by revisiting the two exactly solvable examples considered above.
For the harmonic oscillator (section~\ref{woh}),
the decay of the return probability is given by~(\ref{wroh}),
whereas $\EF\w\approx 2N\o$.
This is in agreement with the above Ansatz, with $B\w=4$,
whereas we had $B=2$ in the two-sided situation.
For the square-well potential (section~\ref{wsquare}),
the decay of the return probability is given by~(\ref{wrnsquare}) at large~$N$,
whereas $\EF\w\approx 2N^2\pi^2/L^2$.
This, too, is in agreement with the above Ansatz, with $B\w=4\e^{-1/2}$,
whereas we had $B=2\e^{-1/2}$ in the two-sided situation.

The relation
\beq
B\w=2B
\label{bwb}
\eeq
between the constants pertaining to the one-sided and two-sided situations,
in the sense of section~\ref{wsym}, is in fact quite general.
This can be shown as follows.
Assume the Ansatz~(\ref{rnas}) holds in the two-sided situation,
in the presence of the symmetrized potential~(\ref{vsym}).
Using~(\ref{ras}), this yields the estimate
\beq
\ln\abs{C_N}\approx\frac{N^2}{4}\left(\ln B-\ln\EF(N)+\frac{3}{2}\right),
\label{lnc}
\eeq
where we have emphasized the dependence of the Fermi energy on the fermion number~$N$.
Consider now the one-sided situation.
First, the correspondence~(\ref{corr}) implies that the Fermi energy
reads approximately
\beq
\EF\w(N)\approx\EF(2N).
\eeq
Second,~(\ref{cforco}) yields
\beq
\ln\abs{C\w_N}\approx\ln\abs{c_N^\odd}\approx\half\ln\abs{C_{2N}}.
\eeq
The last estimate is obtained by expressing that both sectors
equally contribute to the expression~(\ref{cce}) of $C_{2N}$,
to leading order for large $N$.
Combining the two above results with~(\ref{wras}) and~(\ref{lnc}),
we readily obtain the aforementioned relation~(\ref{bwb}).

\section{Discussion}
\label{disc}

We have investigated the quantum return probability
of a system of $N$ non-interacting fermions prepared in the ground state
of a 1D confining potential and submitted to an instantaneous quench
consisting in releasing the trapping potential.
Our main finding is that this return probability
falls off as a power law in the long-time regime,
with a universal exponent which only depends on the fermion number and on the geometry,
equal to $N^2$ when the fermions expand over the full line
and $N(2N+1)$ when the free expansion is constrained to take place over a half-line.
These universal exponents are however not robust with respect to interactions.
For the Calogero-Sutherland model,
an exactly solvable interacting many-particle system on the full line~\cite{Campo2},
the decay exponent is known to be $N(1+\lambda(N-1))$,
where the coupling constant $\lambda$ allows to interpolate between free bosons for $\lambda=0$,
where the exponent is simply $N$,
and hard-core bosons (equivalent to non-interacting fermions in one dimension)
for $\lambda=1$, where the exponent~$N^2$ is recovered.

Table~\ref{expos} presents a comparison of the decay exponent in both geometries
with exponents defined similarly in two other situations,
namely tight-binding lattice fermions launched from a compact configuration,
investigated in our recent work~\cite{KLM},
and non-colliding classical random walkers,
whose survival and return probabilities are derived in~\ref{classical}.
There are both analogies and differences between these three situations.
In all cases, the dependence of the exponent
on the particle number $N$ is a quadratic polynomial with simple coefficients,
and its growth law at large $N$ is twice larger in the half-line geometry than on the full line.
The qualitative differences between continuum and lattice fermions
can be explained in terms of the symmetries of their respective dispersion relations.
In the continuum, the quadratic dispersion relation $E=q^2/2$ has a single minimum at $q=0$,
where the group velocity $v=\dd E/\dd q=q$ vanishes.
On the one-dimensional lattice, the band structure of a tight-binding particle, $E=2\cos q$,
possesses two inequivalent stationary points where the group velocity $v=-2\sin q$ vanishes,
namely $q=0$, as before, and $q=\pi$.
In the long-time regime of the free expansion phase,
both stationary points in momentum space are roughly equally populated~\cite{KLM}.
This band-structure effect has two noticeable consequences on the decay exponent:
it is roughly twice smaller than its counterpart in the continuum
and exhibits a rather unexpected dependence on the parity of $N$.
In the case of non-colliding classical walkers, considered in~\ref{classical},
there are of course no interferences.
Lattice and continuum random walks share the same universal continuum limit,
namely Brownian motion,
and especially the same decay exponents.

\begin{table}[!ht]
\begin{center}
\begin{tabular}{|l|c|c|}
\hline
Model & Exponent (full line) & Exponent (half-line)\\
\hline
Continuum fermions & $N^2$ & $N(2N+1)$\\
\hline
Lattice fermions
$\left\{\!\!\begin{tabular}{l} $N$ even\\$N$ odd\end{tabular}\right.$ & 
\begin{tabular}{l} $\half N^2$\\$\half(N^2+1)$\end{tabular} &
\begin{tabular}{l} $N(N+1)$\\$N^2+N+1$\end{tabular}\\
\hline
Classical walkers & $\quart N(N+1)$ & $\half N(N+1)$\\
\hline
\end{tabular}
\caption{Decay exponent of the return probability for a system of $N$ particles
moving either on the full line or on a half-line.
First row:
Non-interacting fermions in the continuum,
prepared in the ground state of a confining potential and instantaneously released
(body of this work).
Second row:
Non-interacting tight-binding lattice fermions
launched from a compact configuration (Reference~\cite{KLM}).
There, the exponent depends on the parity of the fermion number $N$.
Third row:
Non-colliding classical walkers (\ref{classical} of this work).
There, the exponent governs the power-law fall-off
of the return probability conditioned on survival.}
\label{expos}
\end{center}
\end{table}

The amplitudes of the power-law decay of the quantum return probability
of $N$ continuum fermions in both geometries
have been shown to depend on the confining potential
only through the quantities $C_N$ and $C\w_N$,
expressed in~(\ref{cndef}) and~(\ref{wcndef}) as $N\times N$ determinants
of moments of the one-body bound-state wavefunctions in the potential.
These amplitudes have been worked out explicitly for the harmonic and square-well
potentials (see~(\ref{roh}), (\ref{rsquare}) and~(\ref{wroh}), (\ref{wrsquare})).
The return probabilities have also been demonstrated to simplify
at large fermion numbers, where they obey scaling laws
involving the ratio $N/(\EF t)$,
with $\EF$ being the Fermi energy of the initial state
(see~(\ref{rnas}), (\ref{wrnas})).

Finally,
the investigations pursued in~\cite{KLM} and in the present work
reveal similarities between the dynamics of free fermionic systems and random matrix theory.
This resemblance,
which has also been put forward recently in a static context~\cite{PRAMajumdar},
is essentially due to the effective repulsion felt
both by the eigenvalues of a random matrix and by 1D fermions.
More specifically, Selberg-Mehta integrals stemming from random matrix theory
have been instrumental in deriving most key results.
The outcomes often involve the Barnes $G$-function,
which is also ubiquitous in random matrix theory,
whereas the scaling in~$N^2$ of the exponents listed in table~\ref{expos}
is reminiscent of the scaling of the free energy of matrix models.

\ack

We are grateful to Jean-Marie St\'ephan for very interesting discussions.

\appendix

\section{Classical analogue: non-colliding random walkers}
\label{classical}

A classical analogue of the problem considered in the body of this work
is a collection of $N$ independent random walkers on the line,
or on the half-line, conditioned to never collide.
This system has been studied
by various approaches~\cite{kmg,fisher,hfisher,grabiner,katori}.
The goal of this appendix is to derive in a self-consistent way
many results on the survival and return probabilities,
some of which are already known but scattered in the literature.
The analogies and the differences between the classical and the quantum situations
and between the geometries of the line and of the half-line
are briefly summarized in section~\ref{disc}.
Hereafter we follow the approach initiated by Karlin and McGregor~\cite{kmg},
and pursued by Lindstr\"om~\cite{lind} and Gessel and Viennot~\cite{GV},
yielding to determinantal formulas such as~(\ref{kmgres}).

\subsection{Walkers on the line}

Consider $N$ independent random walkers on the line
starting at time $t=0$ from the positions $\vx=(x_1,\dots,x_N)$, with $x_1<\dots<x_N$.
At any subsequent time $t$,
the probability (or probability density) $P(\vx,\vy,t)$ that
the walkers are at the positions $\vy=(y_1,\dots,y_N)$, with $y_1<\dots<y_N$,
and that their trajectories have not intersected,
is given by the Karlin-McGregor formula~\cite{kmg}
\beq
P(\vx,\vy,t)=\det p(x_i,y_j,t),
\label{kmgres}
\eeq
where $p(x,y,t)$ is the transition kernel (probability or probability density)
for one single walker.
The above determinantal formula holds for several kinds of microscopic realizations
of a random walk, either discrete or continuous.
The only condition is that no particle can jump over another one.
Here are two important examples.

\noindent $\bullet$~For Brownian particles, with diffusion coefficient $D=1/2$,
the transition probability density reads
\beq
p(x,y,t)=\frac{\e^{-(y-x)^2/(2t)}}{\sqrt{2\pi t}}.
\label{pc}
\eeq

\noindent $\bullet$~For particles executing continuous-time random walks
on the lattice of integers,
with jumps to neighboring sites at unit rate,
so that again $D=1/2$, the transition probability reads
\beq
p(x,y,t)=\e^{-t}\,I_{y-x}(t),
\label{pd}
\eeq
where $I_{y-x}$ is the modified Bessel function.
In the continuum limit, i.e., for $t$ large
and distances $\abs{y-x}$ not much greater than the diffusive scale~$\sqrt{t}$,
the discrete expression~(\ref{pd}) becomes the continuous one~(\ref{pc}).

For a long time and for fixed (i.e., bounded) initial positions $x_i$,
and arbitrary final positions $y_j$,
using~(\ref{pc}) allows us to simplify~(\ref{kmgres}) to
\beq
P(\vx,\vy,t)\approx(2\pi t)^{-N/2}\,\det\left(\e^{x_iy_j/t}\right)
\,\exp\!\left(-\frac{1}{2t}\sum_jy_j^2\right).
\eeq
Furthermore, the determinant can be evaluated along the lines of the derivation of~(\ref{detexp}).
We have indeed, to leading order in the regime where all the variables~$x_i$ are small,
\beq
\det(F_j(x_i))\approx\det(F_{j,k-1})\Delta_N(\vx),
\label{fid}
\eeq
where
\beq
F_j(x)=\sum_{k\ge0}F_{j,k}x^k,\qquad F_{j,k}=\frac{F_j^{(k)}(0)}{k!},
\eeq
whereas $\Delta_N(\vx)$ is the Vandermonde determinant of the $x_i$ (see~(\ref{vdm})).
In the present case, $F_j(x)=\e^{xy_j/t}$, and so $F_{j,k}=(y_j/t)^k$.
We are thus left with
\beq
P(\vx,\vy,t)\approx\frac{\Delta_N(\vx)\Delta_N(\vy)}{(2\pi)^{N/2}G(N+1)\,t^{N^2/2}}
\,\exp\!\left(-\frac{1}{2t}\sum_jy_j^2\right).
\label{pasy}
\eeq

The first quantity of interest is the survival probability $S_N(\vx,t)$, i.e.,
the probability that no two trajectories of the $N$ walkers have intersected up to time~$t$.
The behavior of this quantity in the long-time regime
is obtained by integrating~(\ref{pasy})
over the allowed range of final positions ($-\infty<y_1<\dots<y_N<+\infty$).
The result is proportional to a Mehta integral of the form~(\ref{meh1c}).
We thus obtain
\beq
S_N(\vx,t)\approx\frac{\sigma_N\,\Delta_N(\vx)}{G(N+2)\,t^{N(N-1)/4}},
\label{csn}
\eeq
where $\sigma_N$ is given by~(\ref{sigmadef}).
The decay exponent $N(N-1)/4$ of the survival probability
can be found in~\cite{fisher,hfisher,grabiner,katori}.
Reference~\cite{grabiner} also contains the expression of the prefactor.

The second quantity of interest is the return probability $R_N(\vx,t)$, i.e.,
the probability (or probability density)
that the walkers return at (or close to) their initial positions
and that no two trajectories of the $N$ walkers have intersected up to time~$t$.
This quantity can be directly read off from~(\ref{pasy}):
\beq
R_N(\vx,t)\approx\frac{\Delta_N^2(\vx)}{(2\pi)^{N/2}G(N+1)\,t^{N^2/2}}.
\label{crn}
\eeq

The classical analogue of the quantum return probability studied in the body of this work
is the return probability conditioned on survival, i.e.,
\beq
\wt R_N(\vx,t)=\frac{R_N(\vx,t)}{S_N(\vx,t)},
\eeq
whose decay is predicted to be
\beq
\wt R_N(\vx,t)\approx\frac{N!\,\Delta_N(\vx)}{(2\pi)^{N/2}\,\sigma_N\,t^{N(N+1)/4}}.
\label{crsn}
\eeq
The above predictions for both return probabilities seem to be novel.

In the case of continuous-time lattice walks,
if the walkers are launched from any~$N$ consecutive sites, we have
\beq
\Delta_N(\vx)=G(N+1),
\eeq
and so the above expressions read
\beqa
S_N(t)\approx\frac{\sigma_N}{N!\,t^{N(N-1)/4}},
\nonumber\\
R_N(t)\approx\frac{G(N+1)}{(2\pi)^{N/2}\,t^{N^2/2}},
\nonumber\\
\wt R_N(t)\approx\frac{G(N+2)}{(2\pi)^{N/2}\,\sigma_N\,t^{N(N+1)/4}},
\eeqa
where $\sigma_N$ is given by~(\ref{sigmadef}).

Finally, when the number of walkers becomes large,
the above expressions can be further simplified
by means of the expansions~(\ref{lnf}) and~(\ref{lnsf}).
Keeping only leading terms in~$N^2$ in the exponentials, we obtain
\beqa
S_N(t)\sim\left(\frac{N}{2\e^{3/2}t}\right)^{N(N-1)/4},
\nonumber\\
R_N(t)\sim\left(\frac{N}{\e^{3/2}t}\right)^{N^2/2},
\nonumber\\
\wt R_N(t)\sim\left(\frac{2N}{\e^{3/2}t}\right)^{N(N+1)/4}.
\eeqa

\subsection{Walkers on the half-line}

Consider now $N$ independent random walkers on the half-line ($x>0$),
starting at time $t=0$ from the positions $\vx=(x_1,\dots,x_N)$, with $0<x_1<\dots<x_N$.
The probability (or probability density) $P\w(\vx,\vy,t)$ that
the walkers are at the positions $\vy=(y_1,\dots,y_N)$ at time $t$, with $0<y_1<\dots<y_N$,
and that their trajectories have neither intersected nor gone through the origin,
is again given by~(\ref{kmgres}), albeit with
\beq
p\w(x,y,t)=\frac{1}{\sqrt{2\pi t}}\left(\e^{-(y-x)^2/(2t)}-\e^{-(y+x)^2/(2t)}\right)
\label{wpc}
\eeq
for Brownian particles,
and
\beq
p\w(x,y,t)=\e^{-t}\left(I_{y-x}(t)-I_{y+x}(t)\right)
\label{wpd}
\eeq
for particles executing continuous-time random walks on the lattice.
In the continuum limit,
the discrete expression~(\ref{wpd}) again becomes the continuous one~(\ref{wpc}).

For a long time and for fixed (i.e., bounded) initial positions $x_i$,
and arbitrary final positions $y_j$,
the expression~(\ref{kmgres}) simplifies to
\beq
P\w(\vx,\vy,t)\approx\left(\frac{2}{\pi t}\right)^{N/2}\!\det\left(\sinh\frac{x_iy_j}{t}\right)
\exp\!\left(-\frac{1}{2t}\sum_jy_j^2\right).
\eeq
Furthermore, the determinant can be evaluated along the lines of the derivation of~(\ref{wdetexp}).
In the case where all the functions $F_j(x)$ are odd,~(\ref{fid}) becomes
\beq
\det(F_j(x_i))\approx\det(F_{j,k-1})\prod_ix_i\,\Delta_N(\vx^2),
\label{wfid}
\eeq
with
\beq
F_j(x)=\sum_{k\ge0}F_{j,k}x^{2k+1},\qquad F_{j,k}=\frac{F_j^{(2k+1)}(0)}{(2k+1)!},
\eeq
and $\vx^2=(x_1^2,\dots,x_N^2)$.
In the present situation,
we have $F_j(x)=\sinh(xy_j/t)$, and so $F_{j,k}=(y_j/t)^{2k+1}$.
Using~(\ref{pid}), we obtain
\beqa
P\w(\vx,\vy,t)&\approx&2^N\sqrt{\frac{N!}{\pi^NG(2N+2)}}
\frac{\prod_i(x_iy_i)\Delta_N(\vx^2)\Delta_N(\vy^2)}{t^{N(2N+1)/2}}
\nonumber\\
&\times&\exp\!\left(-\frac{1}{2t}\sum_jy_j^2\right).
\label{wpasy}
\eeqa

The survival probability $S\w_N(\vx,t)$ is now the probability
that no trajectory has either crossed the origin or intersected another one up to time $t$.
The behavior of this quantity in the long-time regime
is obtained by integrating~(\ref{wpasy})
over the allowed range of final positions ($0<y_1<\dots<y_N<+\infty$).
The result is proportional to a Mehta integral of the form~(\ref{meh2c}).
We thus obtain
\beq
S\w_N(\vx,t)\approx\frac{2^NG(N+2)\prod_ix_i\,\Delta_N(\vx^2)}
{\sqrt{\pi^N N!G(2N+2)}\,t^{N^2/2}}.
\label{wcsn}
\eeq

The return probability $R\w_N(\vx,t)$
is now the probability (or probability density)
that the walkers return at (or close to) their initial positions
and that no trajectory has either crossed the origin or intersected another one up to time $t$.
This quantity can be directly read off from~(\ref{wpasy}):
\beq
R\w_N(\vx,t)\approx
2^N\sqrt{\frac{N!}{\pi^NG(2N+2)}}
\frac{\prod_ix_i^2\,\Delta_N^2(\vx^2)}{t^{N(2N+1)/2}}.
\label{wcrn}
\eeq
The return probability conditioned on survival scales as
\beq
\wt R\w_N(\vx,t)\approx\frac{\prod_ix_i\,\Delta_N(\vx^2)}{G(N+1)\,t^{N(N+1)/2}}.
\label{wcrsn}
\eeq

In the case of continuous-time lattice walks,
if the walkers are launched from the first $N$ sites
of the half-infinite chain ($x_i=i$), we have
\beq
\prod_ix_i=N!,\qquad\Delta_N(\vx^2)=\sqrt{\frac{G(2N+2)}{2^N N!^3}}.
\eeq
The second equality is equivalent to~(\ref{dpo}).
The above expressions become
\beqa
S\w_N(t)\approx\frac{2^{N/2}G(N+1)}{\pi^{N/2}\,t^{N^2/2}},
\nonumber\\
R\w_N(t)\approx\sqrt{\frac{G(2N+2)}{\pi^N N!}}\frac{1}{t^{N(2N+1)/2}},
\nonumber\\
\wt R\w_N(t)\approx\sqrt{\frac{G(2N+2)}{2^N N!}}\frac{1}{G(N+1)\,t^{N(N+1)/2}}.
\eeqa

Finally, when the number of walkers becomes large,
the above expressions can be further simplified
by means of the expansions~(\ref{lnf}) and~(\ref{lnsf}).
Keeping only leading terms in~$N^2$ in the exponentials, we obtain
\beqa
S\w_N(t)\sim\left(\frac{N}{\e^{3/2}t}\right)^{N^2/2},
\nonumber\\
R\w_N(t)\sim\left(\frac{2N}{\e^{3/2}t}\right)^{N(2N+1)/2},
\nonumber\\
\wt R\w_N(t)\sim\left(\frac{4N}{\e^{3/2}t}\right)^{N(N+1)/2}.
\eeqa

\section{Mehta integrals}
\label{meh}

In this appendix we give the expressions of two of the well-known
Mehta integrals~\cite{mehta}.
These multiple integrals,
which can be derived as limiting values of the Selberg integral
and play a central part in random matrix theory,
are used in the present work at several places.
Reference~\cite{selberg} provides a comprehensive historical overview
of the Selberg and related integrals.

\noindent $\bullet$
First Mehta integral~\cite[Eq.~(17.6.7)]{mehta}:
\beqa
I_1(N,a,\gamma)
&=&
\int_{-\infty}^\infty\etc\int_{-\infty}^\infty
\prod_{n=1}^N\left(\dd x_n\,\e^{-ax_n^2}\right)
\abs{\Delta_N(\vx)}^{2\gamma}
\nonumber\\
&=&
\frac{(2\pi)^{N/2}}{(2a)^{N(1+(N-1)\gamma)/2}}
\prod_{j=1}^N\frac{\Gamma(1+j\gamma)}{\Gamma(1+\gamma)}.
\label{meh1}
\eeqa
We have in particular
\beq
I_1(N,a,1)=\frac{(2\pi)^{N/2}G(N+2)}{(2a)^{N^2/2}}
\label{meh1q}
\eeq
for $\gamma=1$,
where $G$ denotes the Barnes $G$-function (see~\ref{barn}),
and
\beq
I_1(N,a,1/2)=\frac{(2\pi)^{N/2}\,\sigma_N}{(2a)^{N(N+1)/4}},
\label{meh1c}
\eeq
for $\gamma=1/2$,
with
\beq
\sigma_N=\frac{\sqrt{\Gamma(\frac{N}{2}+1)G(N+2)}}{2^{N(N-3)/4}\pi^{N/4}}.
\label{sigmadef}
\eeq

\noindent $\bullet$
Second Mehta integral~\cite[Eq.~(17.6.6)]{mehta}:
\beqa
I_2(N,a,\beta,\gamma)
&=&
\int_{-\infty}^\infty\etc\int_{-\infty}^\infty
\prod_{n=1}^N\left(\dd x_n\,\abs{x_n}^{2\beta-1}\e^{-ax_n^2}\right)
\abs{\Delta_N(\vx^2)}^{2\gamma}
\nonumber\\
&=&
\frac{1}{a^{N(\beta+(N-1)\gamma)}}
\prod_{j=1}^N\frac{\Gamma(1+j\gamma)\Gamma(\beta+(j-1)\gamma)}{\Gamma(1+\gamma)}.
\label{meh2}
\eeqa
We have in particular
\beq
I_2(N,a,3/2,1)=\frac{\sqrt{\pi^NN!G(2N+2)}}{(2a)^{N(2N+1)/2}}
\label{meh2q}
\eeq
for $\beta=3/2$ and $\gamma=1$,
and
\beq
I_2(N,a,1,1/2)=\frac{2^NG(N+2)}{(2a)^{N(N+1)/2}}
\label{meh2c}
\eeq
for $\beta=1$ and $\gamma=1/2$.

\section{The Barnes $G$-function and related identities}
\label{barn}

Let us begin with a brief summary of the main properties of the Barnes $G$-function
(see~\cite{WW}).
The Euler $\Gamma$-function and the Barnes $G$-function
are meromorphic functions of the complex variable $z$
obeying the recursion relations
\beq
\Gamma(z+1)=z\Gamma(z),\qquad G(z+1)=\Gamma(z)G(z).
\eeq
When $z$ is a positive integer,
the $\Gamma$-function becomes the usual factorial:
\beq
\Gamma(n+1)=n!,
\eeq
whereas the $G$-function becomes the `superfactorial':
\beq
G(n+2)=\prod_{k=1}^n k!=\prod_{\ell=1}^n\ell^{n+1-\ell}=\prod_{1\le i<j\le n+1}(j-i).
\eeq

The $\Gamma$ and $G$-functions
have the following asymptotic expan\-sions as $z\to+\infty$:
\beqa
\ln\Gamma(z+1)
&=&\left(z+\half\right)\ln z-z+\half\ln(2\pi)+\frac{1}{12z}+\cdots,
\label{lnf}
\\
\ln G(z+2)
&=&\left(\frac{z^2}{2}+z+\frac{5}{12}\right)\ln z-\frac{3z^2}{4}-z
\nonumber\\
&+&\frac{z+1}{2}\ln(2\pi)+\zeta'(-1)+\cdots,
\label{lnsf}
\eeqa
where $\zeta'(-1)=-0.165421$ ($\zeta$ being the Riemann $\zeta$-function).

The following products can be expressed in terms of values of the $G$-function:
\beqa
\prod_{k=1}^n(n-k)!=G(n+1),
\\
\prod_{k=1}^n(n+k)!=\frac{G(2n+2)}{G(n+2)},
\\
\prod_{k=1}^n(n+k-1)!=\frac{G(2n+1)}{G(n+1)},
\\
\prod_{k=1}^n(2k)!=\sqrt{2^n n! G(2n+2)},
\\
\prod_{k=1}^n(2k-1)!=\sqrt{\frac{G(2n+2)}{2^n n!}}.
\label{pid}
\eeqa

The above identities allow us to evaluate in closed form
the determinants $d_p^\even$ and $d_p^\odd$
introduced in~(\ref{dpedef}) and~(\ref{dpodef}).
We have indeed
\beqa
d_p^\even&=&(-1)^{p(p-1)/2}\det((2k+1)^{-2l})_{0\le k,l\le p-1}
\nonumber\\
&=&\prod_{1\le k<l\le p}\left(\frac{1}{(2k-1)^2}-\frac{1}{(2l-1)^2}\right)
\nonumber\\
&=&\prod_{1\le k<l\le p}\frac{4(l-k)(l+k-1)}{(2k-1)^2(2l-1)^2}
\nonumber\\
&=&\frac{2^{p(p-1)}}{((2p-1)!!)^{2(p-1)}}\prod_{k=1}^p\frac{(p-k)!(p+k-1)!}{(2k-1)!}
\nonumber\\
&=&2^{p(3p-2)}\left(\frac{p!}{(2p)!}\right)^{2p-1}\sqrt{\frac{G(2p+2)}{2^pp!}}
\label{dpe}
\eeqa
and
\beqa
d_p^\odd
&=&(-1)^{p(p-1)/2}\det(k^{-2(l-1)})_{1\le k,l\le p}
\nonumber\\
&=&\prod_{1\le k<l\le p}\left(\frac{1}{k^2}-\frac{1}{l^2}\right)
\nonumber\\
&=&\prod_{1\le k<l\le p}\frac{(l-k)(l+k)}{k^2l^2}
\nonumber\\
&=&\frac{1}{p!^{2(p-1)}}\prod_{k=1}^p\frac{(p-k)!(p+k)!}{(2k)!}
\nonumber\\
&=&\frac{1}{p!^{2p-1}}\sqrt{\frac{G(2p+2)}{2^pp!}}.
\label{dpo}
\eeqa

\end{document}